\documentclass{ws-procs9x6}            
\begin{document}
\title{
Spectroscopy of $a_1$ mesons from lattice QCD with the truncated overlap fermions
}

\author{
Masayuki~Wakayama$^{1,2,3,\ast}$, 
Yuko~Murakami$^{4}$, 
Atsushi~Nakamura$^{5,6,3}$, 
Motoo~Sekiguchi$^{7}$ and 
Hiroaki~Wada$^{7}$
}
\address{
$^1$Center for Extreme Nuclear Matters (CENuM), Korea University,\\
Seoul 02841, Republic of Korea\\
$^2$Department of Physics, Pukyong National University (PKNU),\\
Busan 48513, Republic of Korea\\
$^3$Research Center for Nuclear Physics (RCNP), Osaka University,\\
Ibaraki, Osaka 567-0047, Japan\\
$^4$Research and Development Laboratory, Seikow Chemical Engineering \& Machinery, LTD,
Akashi 674-0093, Japan\\
$^5$School of Biomedicine, Far Eastern Federal University, 690950 Vladivostok, Russia\\
$^6$Theoretical Research Division, Nishina Center, RIKEN, Wako 351-0198, Japan\\
$^7$School of Science and Engineering, Kokushikan University, Tokyo 154-8515, Japan
$^{\ast}$E-mail: wakayama@rcnp.osaka-u.ac.jp
}

\begin{abstract}
We study the ground state and next radial excitation of the $a_1$ mesons 
from a quenched lattice QCD simulation with the truncated overlap fermions formalism 
based on domain wall fermions. 
Our results are consistent with the experimental values for $a_1$(1260) and $a_1$(1640).
\end{abstract}


\bodymatter

\section{Introduction}\label{Intro}
In hadron physics, 
determining the properties of excited light mesons will provide 
important information on the chiral dynamics of quantum chromodynamics (QCD). 
We study the structures of $a_1$ mesons determined using 
the truncated overlap fermion (TOF) formalism 
by Bori\c{c}i~\cite{Borici:1999zw} 
based on the domain wall fermion formalism~\cite{Kaplan:1992bt,Furman:1994ky}. 
The TOF formalism is classified into lattice chiral fermions
~\cite{Kaplan:1992bt,Furman:1994ky,Narayanan:1993ss,Brower:2004xi}, 
and exhibits good chiral symmetry. 
Our objective is to reveal 
the relationship between the nature of the $a_1$ meson associated with the chiral partner of the $\rho$ meson 
and dynamical chiral symmetry breaking, 
as for the $\pi$ and the chiral partner of the $\sigma$ meson~\cite{Nambu}. 
In the conventional constituent quark model, the $\sigma$ meson and
the $a_1$ meson are assigned excited states. 

Lattice simulations of $a_1$ mesons have been previously conducted. 
Wingate {\it et al.} were the first to measure the mass of the $a_1$ meson using two-flavor lattice QCD~\cite{Wingate:1995hy}. 
Their result agrees with the experimental value for $a_1$(1260). 
A decade later, a study of the $a_1$ meson 
was performed with the L${\rm \ddot{u}}$scher--Weisz gauge action and 
the chirally improved Dirac operator in the quenched approximation~\cite{Gattringer:2008}. 
Although the ground state of the $a_1$ meson was improved 
by using various interpolators including derivative quark sources in the simulations, 
the obtained mass of the ground state is close to $a_1$(1420), instead of $a_1$(1260). 
Moreover, the mass of the first excited state of $a_1$ was observed to be above 2~GeV. 

Recently, Prelovsek {\it et al.} presented 
results for the mass of the $a_1$ meson and its coupling constant~\cite{Prelovsek:2011im}. 
They performed a simulation for a full QCD lattice with clover-improved Wilson quarks. 
This work was continued in Ref.~\cite{Lang:2014tia}, in which 
they extracted the resonance mass of the ground state of the $a_1$ meson $m_{a_1} = 1.435(53)(_{-109}^{+0})$~GeV 
and the coupling $g_{a_1\pi\rho}=1.71(39)$~GeV 
by simulating the corresponding scattering channel $\pi$$\rho$. 
Their obtained value of the $a_1$ meson mass is higher than the experimental result of $a_1$(1260)~\cite{COMPASS:2010,PDG2018}.
In our previous work, 
we investigated the mass of the ground state of the $a_1$ meson by a quenched lattice QCD using TOF~\cite{Wakayama:2019crb}. 
Our result is consistent with the experimental value of $a_1$(1260).

\section{Truncated overlap fermions}\label{TOF}

The TOF~\cite{Borici:1999zw} are defined by 
\begin{eqnarray}
D_{TOF} &=&  \epsilon^{\dag} P^{\dag} D_{PV}^{-1}D_{DWF} P \epsilon \ , \label{TOF_def} \\
\epsilon_{x_5}   \ = \  \delta_{1,x_5} \ , \ \ \ 
 P_{x_5y_5} &=& P_{L} \delta_{x_5,y_5} + P_{R} \delta_{x_5+1,y_5} + P_{R} \delta_{x_5,N_5} \delta_{y_5,1}  \  ,  
\end{eqnarray}
where the five dimensional projection operator $P$ is constructed 
from the four-dimensional projection operators $P_ {R/L}=(1\pm\gamma_5)/2$. 
The indexes $x_5$ and $y_5$ represent the fifth-dimensional lattice sites, which are defined in $x_5,y_5 \in [1,N_5]$. 
The domain wall fermion operator $D_{DWF}$~\cite{Kaplan:1992bt,Furman:1994ky} is defined by 
\begin{eqnarray}
 D_{DWF \, x_5 y_5}(x,y) 
     &=& D_{WF} (x,y) \delta_{x_5,y_5}
            - ( P_L \delta_{x_5+1,y_5}  + P_R\delta_{x_5-1,y_5} ) \delta_{x,y} \nonumber \\
      &+& \delta_{x,y} \delta_{x_5,y_5} + m_f (P_R \delta_{x_5,1}\delta_{y_5,N_5} + P_L \delta_{x_5,N_5}\delta_{y_5,1} ) \delta_{x,y} \ , \ \ \ \ \ 
\end{eqnarray}
where $D_{WF}$ is the Wilson fermion operator, 
\begin{eqnarray}
 D_{WF}(x,y) &=& (4-M_5) \delta_{x,y} 
                       - \frac{1}{2} {\textstyle \sum_{\mu=\pm 1}^{\pm 4}} (1-\gamma_\mu)U_{\mu}(x)\delta_{x+\hat{\mu},y} \ .
\end{eqnarray}
The parameters of TOF are $m_f$ and $M_5$, 
which correspond to the bare quark mass and the height of the domain wall, respectively. 
The Pauli--Villars matrix $D_{PV}$ is given by $D_{PV}=D_{DWF}(m_f=1)$. 
In the $N_5\to\infty$ limit, the lattice chiral symmetry is exactly reproduced in TOF.

\section{Simulation setup and Lattice QCD results}\label{setup}

In this work, we simulate the spectroscopy of $a_1$ mesons on $8^3\times24$ quenched lattice with the plaquette gauge action with $\beta=5.7$. 
Gauge configurations are generated with the pseudo-heat-bath method. 
After 20000 thermalization iterations, we start to save gauge configurations every 1000 sweeps. 
The propagators of the $\pi$, $\rho$ and $a_1$ mesons are calculated with TOF. 
The fermion parameters are set to $N_5=32$, $M_5=1.65$ and $m_f a=0.04$--0.08.

\begin{figure}
\begin{center}
\includegraphics[width=2.05in]{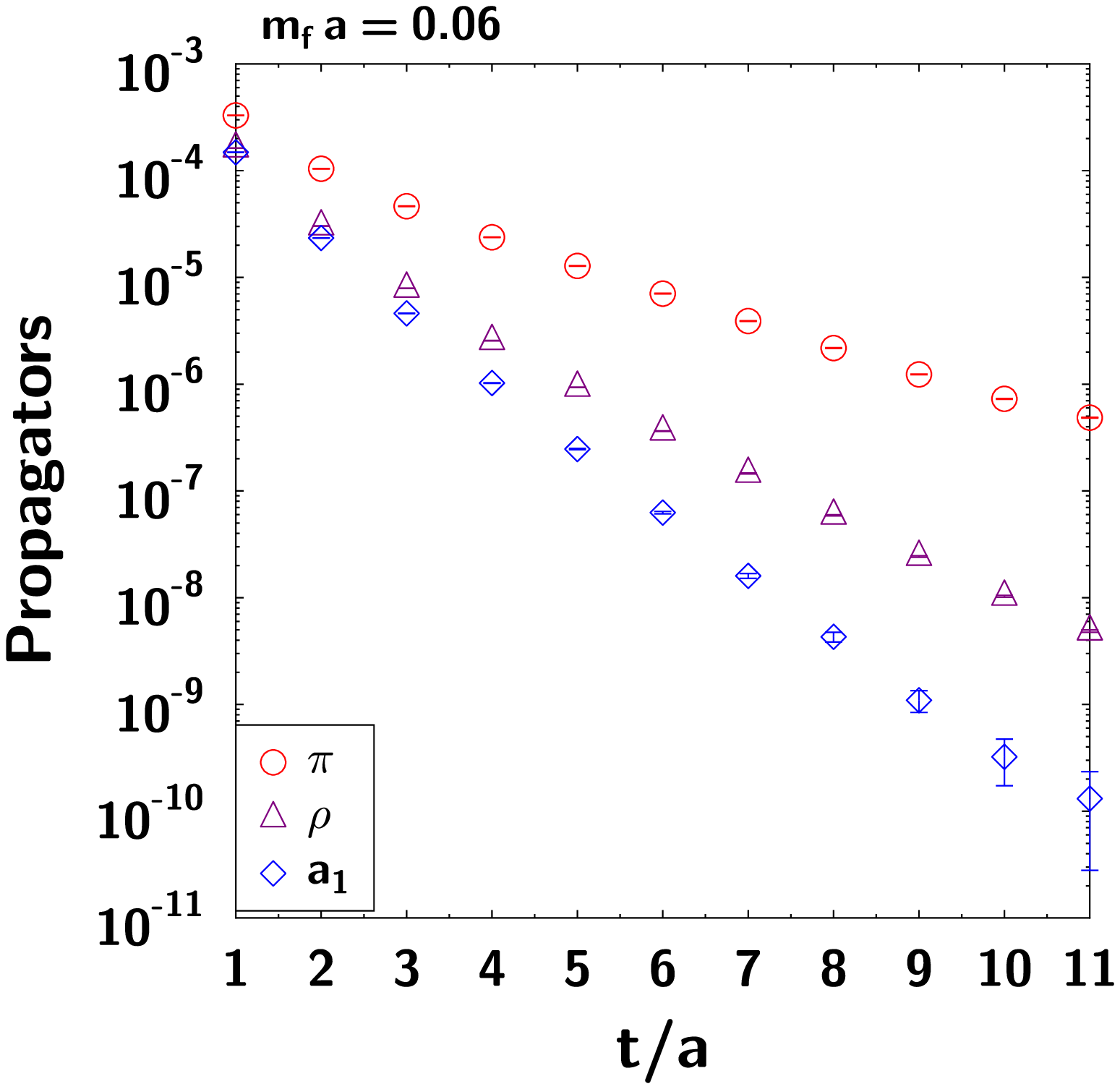} 
\includegraphics[width=2.05in]{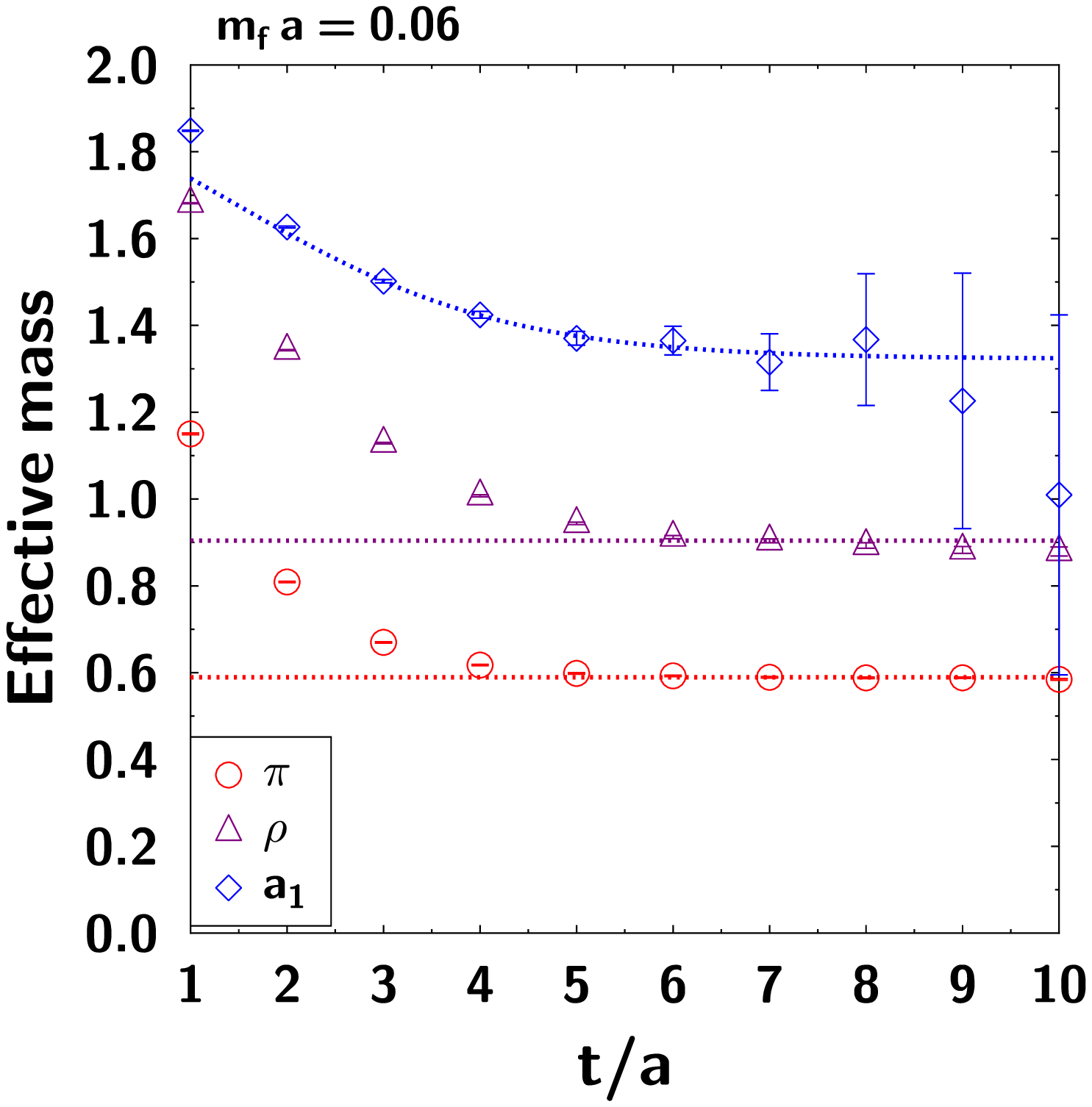}
\end{center}
\caption{
Time dependences of the propagators (left) and the effective masses (right) for $m_fa=0.06$. 
In the right figure, 
the dotted lines and dotted curve represent the single-pole and double-pole fitting results, respectively. 
}
\label{fig1_propa}
\end{figure}

We calculate the meson propagators and their effective masses (see Fig.~\ref{fig1_propa}). 
We estimate the statistical errors using the jackknife method. 
By performing a single- or double-pole fit to the effective masses, 
we obtain the meson masses as listed in Table~\ref{tbl1_para}.

\begin{table}
\tbl{Masses of the $\pi$ and $\rho$ mesons, mass ratio $m_{\pi}/m_{\rho}$ and number of configurations for each quark mass. }
{\begin{tabular}{@{}cccccc@{}}\toprule
\ $m_f a$ \ & \ $m_{\pi}a$ \ & \ $m_{\rho}a$ \ & \  $m_{\pi}/m_{\rho}$ \ & \ Confs. \ 
\\\colrule
0.08 & 0.6668(7) & 0.9496(18) & 0.702(2) & 3000 \\
0.07 & 0.6283(7) & 0.9249(21) & 0.679(2) & 3000 \\
0.06 & 0.5895(8) & 0.9042(24) & 0.652(3) & 3000 \\
0.05 & 0.5478(8) & 0.8816(27) & 0.621(3) & 3600 \\
0.04 & 0.5028(6) & 0.8614(24) & 0.584(2) & 7864 \\\botrule
\end{tabular}}
\label{tbl1_para}
\end{table}

\begin{figure}
\begin{center}
\includegraphics[width=3.42in]{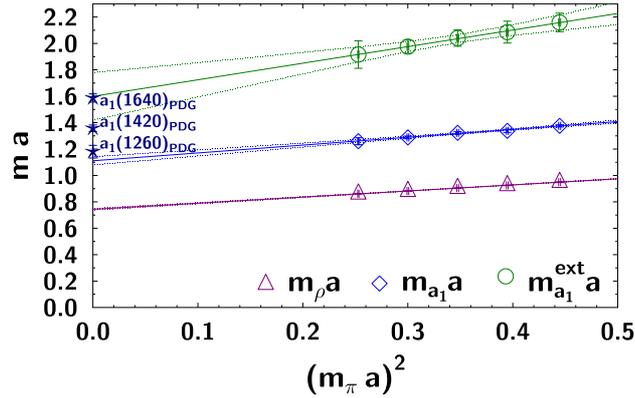}
\end{center}
\caption{
Squared pion mass dependence of the meson masses. }
\label{fig2_mass}
\end{figure}

Fig.~\ref{fig2_mass} shows the quark mass dependence of the meson masses. 
We linearly extrapolate the meson masses to the chiral limit, $(m_{\pi}a)^2= 0$. 
By tuning the $\rho$ meson mass in the chiral limit to $m_{\rho} = 775$~MeV, 
we obtain a lattice spacing of $a = 0.1893(15)$~fm. 
Note that calculations for TOF have a residual mass such as $m_fa=-m_{\rm res} a$ in the chiral limit 
due to the finite $N_5$ effect. 
In our calculation with $N_5=32$, 
we obtain $m_{\rm res} a = 1.29(4) \times 10^{-2}$, that is $m_{\rm res} = 13.4(5)$~MeV, 
which is negligible. 
We estimate the masses of the ground state and first excited states of the $a_1$ mesons to be 
$m_{a_1} = 1158(42)$~MeV and 
$m_{a_1}^{\rm ext} = 1667(202)$~MeV. 
Our results are consistent with the experimental values of $a_1$(1260) and $a_1$(1640), 
$m_{a_1(1260)}$=1230(40)~MeV and $m_{a_1(1640)}$=1654(19)~MeV~\cite{PDG2018}.

\section{Conclusion}\label{conc}
We have investigated the masses of the ground and first excited states of $a_1$ mesons from a quenched lattice QCD 
with the TOF action. 
We have obtained the masses of the $a_1$ mesons to be 1158(42)~MeV and 1667(202)~MeV, 
which are in good agreement with the experimental values of $a_1$(1260) and $a_1$(1640). 
Since our simulation was performed in the quenched approximation and the $\bar{q}q$ source and sink, 
in which virtual intermediate states such as $\bar q \bar q qq$ states are highly suppressed, 
our results suggest that 
$a_1$(1260) and $a_1$(1640) are the simple $\bar{q}q$ states 
whereas $a_1$(1420) may have a more complicated structure than the $\bar q q$ state.

\section*{Acknowledgments}
This work was supported by RSF grant 15-12-20008 
and the National Research Foundation of Korea (NRF) grant funded by the Korean government (MSIT) (No.~2018R1A5A1025563). 
The simulations were performed on the supercomputer system SX-ACE at RCNP and the Cybermedia Center, Osaka University, 
and were conducted using the Fujitsu PRIMEHPC FX10 System (Oakleaf-FX, Oakbridge-FX) at the Information Technology Center, The University of Tokyo. 
This work was supported by ``Joint Usage/Research Center for Interdisciplinary Large-scale Information Infrastructures'' in Japan (Project ID: jh180053-NAJ and jh190048-NAH).



\end{document}